\documentstyle[preprint,aps,psfig, epsfig,12pt]{revtex}
\begin{document}
\draft
\preprint{\vbox{\hbox{JLAB-THY-98-51}
}}

\def\beq{\begin{equation}}
\def\eeq{\end{equation}}
\title{The Goldberger-Treiman Discrepancy in SU(3)}
%\vspace*{1.cm}\\
\author{Jos\'e L. Goity$^{1,2}$, 
Randy Lewis$^{1,3}$, 
Martin Schvellinger$^{1,2,4}$
and Longzhe Zhang$^{1,2}$}

\address{$^1$Jefferson Lab,
         12000 Jefferson Avenue, Newport News, VA, 23606, U.S.A.}
\address{$^2$Department of Physics, Hampton University, Hampton,
         VA, 23668, U.S.A.}
\address{$^3$Department of Physics, University of Regina, Regina,
         SK, S4S 0A2, Canada}
\address{$^4$Department of Physics, Universidad Nacional de La Plata, 
     C.C. 67,    1900 La Plata, Argentina}

%\date{\today}
\maketitle
%\thispagestyle{empty}
%\begin{abstract}

\vspace{5mm}

%\flushright{JLAB-THY-98-}
%\flushright{LA PLATA-TH-98-??}

\begin{abstract}

The Goldberger-Treiman discrepancy in SU(3) is analyzed in the framework of  
heavy baryon chiral perturbation theory (HBChPT).  
It is shown that the discrepancy at leading order is entirely given 
by counterterms from the ${\cal O}(p^3)$ Lagrangian, and that the first
subleading corrections are suppressed by two powers in the HBChPT expansion. 
These subleading corrections include meson-loop contributions as well as
counterterms from the ${\cal O}(p^5)$ Lagrangian.
Some one-loop contributions are calculated and found to be small.
Using the three discrepancies ($\pi{N}N$, $KN\Lambda$ and $KN\Sigma$)
which can be extracted from existing experimental data, we find that
the HBChPT calculation favors the smaller $g_{\pi{N}N}$
values obtained in recent partial wave analyses.
% \approx 13.0$ rather than 13.6.
\end{abstract}
\vspace*{10mm}
\noindent goity@jlab.org\\
randy.lewis@uregina.ca\\
martin@venus.fisica.unlp.edu.ar\\
lzhang@jlab.org
\thispagestyle{empty}
%\pacs{{\tt$\backslash$\string pacs\{\}}}
\newpage
\setcounter{page}{1}

\section{Introduction}
 
The Goldberger-Treiman relation (GTR)\cite{GT},
obtained from matrix elements of the divergence of axial currents between 
spin 1/2 baryons, is an important indicator of explicit chiral symmetry 
breaking by the quark masses. It interrelates baryon masses, 
axial vector couplings, the baryon-pseudoscalar meson (Goldstone boson
$\equiv$ GB) couplings and the GB decay constants.
Explicit chiral symmetry breaking leads to a departure from the GTR (defined
below) which is called the Goldberger-Treiman discrepancy (GTD).
 
The GTD has been repeatedly discussed over time\cite{Dominguez}
and for several reasons there were difficulties in arriving at a 
clear understanding.  On one hand, there was no available
effective theory with a systematic expansion
to address the problem, and on the other hand
the experimental values of the baryon-GB couplings were too
poorly known. In recent years,
progress has been made on both fronts.
There is now a baryon chiral effective theory that permits a consistent 
expansion of the discrepancy\cite{Gasser,HBChPT,HUGS,Stern}.
There has also been
progress in the determinations of the baryon-GB couplings
that are the main source of uncertainty in the phenomenological extraction
of the discrepancies. In fact, 
the current knowledge of the couplings $g_{\pi NN}$, $g_{K N \Lambda}$
and, to a lesser extent, $g_{K N \Sigma}$ is good enough to justify a new look
at the GTD in SU(3).  In this work we study the GTD in the light of
heavy baryon chiral perturbation theory (HBChPT)\cite{HBChPT,HUGS}.

Let us first briefly review the derivation  of the GTR\cite{Dashen} and the definition of the GTD.
We consider the matrix elements of the octet axial
current $A_\mu^a=\frac{1}{2}\bar{q}(x)\gamma_\mu \gamma_5 \, \lambda^a\, q(x)$ 
(the Gell-Mann matrices are normalized to
${\rm Tr}(\lambda^a \lambda^b)=2 \delta^{ab}$)
between states of the baryon octet:
\begin{eqnarray}
 \left< b, \, p_b\mid A_\mu^c\mid a, \, p_a\right>=
 \bar{U}_b(p_b)\left[ \frac{1}{2}\,\gamma_\mu g_A^{abc}(q^2)-q_\mu g_2^{abc}(q^2)\right]
 \gamma_5  U_a(p_a) \,\,\, ,
\label{1}
\end{eqnarray}
where $a,b,c=1,...,8$ and $q=p_b-p_a$ is the momentum transfer between
baryons $a$ and $b$. 
{}From Eq. (\ref{1}), the matrix elements
of the divergence of the axial currents become 
\begin{eqnarray}
 \left< b, \, p_b\mid \partial^\mu A_\mu^c\mid a, \, p_a \right>=
 i \bar{U}_b(p_b)\left[ -\frac{1}{2}\,(M_a+M_b) g_A^{abc}(q^2)+q^2 g_2^{abc}(q^2)
 \right] \gamma_5 U_a(p_a) \,\,\, ,
 \label{2}
\end{eqnarray}
where $M_a$ is a baryon mass. 
Crucial to the derivation of
the GTR is the GB pole contribution represented in Fig. 1. 
To explicitly expose the pole term, the matrix element in Eq. (\ref{2})
can be rewritten as
 \begin{eqnarray}
 \left< b, \, p_b\mid \partial^\mu A_\mu^c\mid a, \, p_a \right>=
 i \bar{U}_b(p_b)\frac{N^{abc}(q^2)}{q^2-m_c^2+i\epsilon}
 \gamma_5 U_a(p_a) \,\,\, ,
 \label{3}
\end{eqnarray}
where $N^{abc}(q^2)=g_{cab}(q^2)P^c(q^2)+(q^2-m_c^2)\delta^{abc}(q^2)$,
$m_c$ is a GB mass, and $g_{cab}(q^2)$ the baryon-GB form factor, defined such that
in the physical basis of the Gell-Mann matrices
$g_{3,6+i7, 6-i7}(M_\pi^2)$ is equal to $g_{\pi^0 nn}$, etc.
$P^c(q^2)$ represent the couplings of the pseudoscalar
currents to the GB's, given in the chiral limit by
$P^c= m_c^2 F_c$ ($F_c$ is the   decay constant, where $F_\pi=92.42$ MeV);
the $q^2$ dependence of $P^c(q^2)$ starts at ${\cal O}(p^4)$ and is henceforth 
disregarded. 
Finally, $\delta^{abc}(q^2)$ denotes contributions
not involving the GB pole, and it starts as a quantity of  ${\cal O}(p^2)$.
 This separation of pole and
non-pole contributions is not unique (the off-shell functions separately
are not observables); for instance, up to higher order terms in $q^2$,
we can choose to remove the $q^2$ dependence in $g_{cab}(q^2)$ around the
point $q^2=m_c^2$ by a simple redefinition of $\delta ^{abc}$.

 In the chiral limit $\partial^\mu A_\mu^c=0$, and at
$q^2=0$ Eq. (\ref{2}) gives:
\begin{equation}
  M g_A^{abc}(0)=\lim_{q^2\to 0}q^2 g_2^{abc}(q^2) = F_c\;g_{cab}(0) \,\,\, ,
\label{4}
\end{equation}
which is the general form of the GTR. Here $M$ is the common
octet baryon mass in the chiral limit.
In the real world, chiral symmetry is explicitly broken by the quark masses
and the GB's become massive. 
In this case, Eqs. (\ref{2}) and (\ref{3}) lead to
\begin{equation}
 {m_c}^2 \, g_2^{abc}(m_c^2)=\lim_{q^2\to m_c^2}
 \frac{g_{cab}(q^2) \, P^c(q^2)}{q^2-{m_c}^2+i \epsilon} \,\,\, .
\label{5}
\end{equation}
In order to define the GTD it is also convenient
to take the limit $q^2\to 0$ which gives
\begin{equation}
 (M_a+M_b) g_A^{abc}(0)=\frac{1}{m_c^2}g_{cab}(0)\,
 P^c(0)-\delta^{abc}(0) \,\,\, .
\label{6}
\end{equation}
The discrepancy $\Delta^{abc}$ is then defined by: 
\begin{equation}
 (M_a+M_b)\, g_A^{abc}(0)=\frac{(1-\Delta^{abc})}{m_c^2} g_{cab}(m_c^2)
 P^c(m_c^2) \,\,\, .
\label{7}
\end{equation}
Notice that while the GTR, Eq. (4), is defined at $q^2=0$,  the
GTD in Eq. (\ref{7}) is given at $q^2={m_c}^2$ because only at that point
is the coupling $g_{cab}$ unambiguously determined. 
At leading order in the quark masses, the GTD can then be expressed
 as follows:
\begin{equation}
 \Delta^{abc}= m_c^2 \frac{\partial}{\partial q^2}
\log N^{abc}(q^2)\mid_{q^2=m_c^2}
\label{8}.
\end{equation}

\section{Tree Level Contributions}

Throughout we are going to use standard definitions, namely:
\begin{eqnarray} 
u&\equiv & \exp\left(-i\frac{\pi^a\lambda^a}{2 F_0}\right), \\
\chi&\equiv & 2 B_0\,(s+i \, p) \,\,\, , \\
\chi_\pm &\equiv & u^\dagger \chi u^\dagger\pm u \chi^\dagger u \,\,\, , \\
\omega^\mu&\equiv &\frac{i}{2} (u^+ \partial^\mu u-u \partial^\mu u^+)  ,  \\
S_v^\mu &\equiv & \frac{i}{2}\gamma_5 \sigma^{\mu\nu} v_\nu \,\,\, .
\label{chi}
\end{eqnarray}
The HBChPT Lagrangian is ordered in powers of momenta and GB 
masses, which are small
compared to both the chiral scale and the baryon masses,
\begin{equation}
   {\cal L} = {\cal L}^{(1)} + {\cal L}^{(2)} + {\cal L}^{(3)} + \ldots~.
\end{equation}
Although the Lagrangian is written as a single expansion, it will
be useful to keep track of
the chiral and $1/M$ suppression factors separately.
As will be demonstrated explicitly below,
leading order (LO) contributions to the GTD appear within
${\cal L}^{(3)}$.  Subleading contributions
are suppressed by at least two suppression factors, so we 
will refer to any contribution at the
order of ${\cal L}^{(5)}$ as a next-to-leading
order (NLO) contribution.

 The tree level contributions to the GTD stem from contact terms in the
effective Lagrangian that can contribute to $\delta^{abc}$, and also 
from terms that
can give a $q^2$ dependence to $g_{cab}$. First we notice that
in HBChPT such terms must contain the spin operator $S_v^\mu$ that results from
the non-relativistic reduction of the baryon pseudoscalar density. 
There are two types of terms which contribute to the GTD.
The first type must contain the pseudoscalar source $\chi_{-}$. 
The second type must contain
monomials such as $[{\cal D}^\mu,\,[{\cal D}_\mu,\,\omega_\nu]]$ and 
$[{\cal D}^\nu,\,[{\cal D}_\mu,\,\omega^\mu]]$ between the baryon field
operators (here  ${\cal D}^\mu$ is the chiral covariant derivative).
However, upon using the classical equations of motion
satisfied by the GB fields at ${\cal O}(p^2)$, it turns out that terms of the 
second type can 
be recast into terms among which there are terms of the first type. 
In this way, one moves the explicit 
$q^2$ dependence from $g_{cab}$ to contact terms, some of which contribute to
$\delta^{abc}$.  Such reduction of terms has been implemented for
${\cal L}^{(1)} + {\cal L}^{(2)} + {\cal L}^{(3)}$,
for instance in Ref. \cite{Ecker}, and in the relativistic effective 
Lagrangian as well\cite{Gasser}.  Some terms in ${\cal L}^{(3)}$  
whose coefficients are determined by  reparametrization invariance \cite{Ecker}
 may seem at first glance to give a $q^2$ dependence 
to $g_{cab}$, but a careful calculation shows that this is   not so.
  
Since $\chi_-$ is ${\cal O}(p^2)$, and since a factor of the spin operator
$S_v^\mu$ is needed, the LO tree contributions 
to the GTD
must come from ${\cal L}^{(3)}$.
One can further argue that there are no contributions from
the even-order Lagrangians, ${\cal L}^{(2 n)}$. The reason is that
an even number of derivatives would require factors in the monomial
of the form $v.\nabla$ which, when acting on the baryon field, are 
in effect replaced by $\nabla^2/2M$; the other possibility would be  
factors of  $v.S_v$ that vanish. 

In the case of SU(2), the Lagrangian ${\cal L}^{(3)}$ has
been given by Ecker and Moj\v{z}i\v{s}\cite{Ecker,bugfix}.
There are only two terms in ${\cal L}^{(3)}$ that
are of interest to us, namely the terms ${\cal O}_{19}$ and ${\cal O}_{20}$
given in Refs. \cite{Ecker,Lewis}.
In the scheme used by Ecker
and Moj\v{z}i\v{s} these are finite counterterms. 
We note that although ${\cal O}_{17}$ and
${\cal O}_{18}$ do contribute to $g_{cab}$ and to $g_A^{abc}$, they
  are such that no contribution to the  GTD
results, as  noticed  in Ref. \cite{Lewis}. 
 In SU(3) there are instead
three ${\cal L}^{(3)}$ terms that are of interest to us, namely,
\begin{eqnarray}
{\cal {L}}^{(3)}_{GTD} &=& iF_{19} {\rm Tr} ({\bar{B}} S_v^\mu
[\nabla_\mu\chi_{-}, \, B]) \nonumber \\
&+& iD_{19} {\rm Tr} ({\bar{B}} S_v^\mu \{\nabla_\mu\chi_{-}, \, B\}) \nonumber \\
&+& ib_{20} {\rm Tr} ({\bar{B}} S_v^\mu B) \, {\rm Tr} (\partial_\mu \,\chi_{-}).
\label{L3}
\end{eqnarray}
The NLO contributions come from ${\cal {L}}^{(5)}_{GTD}$ and will not be
displayed here. There are, for instance, terms
quadratic in the quark masses such as  ${\rm Tr} ({\bar{B}} S_v^\mu
\chi_+[\nabla_\mu\chi_{-}, \, B])$ and others. 

 The contribution to 
$\delta^{abc}$ from  ${\cal {L}}^{(3)}_{GTD}$  is given by
\begin{equation}
 \frac{\delta^{abc}_{CT}}{4MB_0}= 2 s_0[i F_{19} f^{abc}+D_{19} d^{abc}]
 +  d^{cde} s^d [i F_{19} f^{bea}+D_{19} d^{abe}]+s^c
 [ \frac{2}{3} D_{19}+b_{20}] \delta^{ab}  \,\,\, ,
\label{deltaCT}
\end{equation}
where
\begin{eqnarray}
s_0&=&\frac{1}{3} (m_u+m_d+m_s) \,\,\, , \\
s^a&=& \delta_{a3} 
(m_u-m_d) 
-\frac{1}{\sqrt{3}} \delta_{a8} (2 m_s-m_u-m_d).
\label{s}
\end{eqnarray}
In deriving Eq. (\ref{deltaCT}) from Eq. (\ref{L3}) we used the Ward identity:
\begin{equation}
\partial^\mu A_\mu^a=2 s_0 \frac{\delta {\cal L}}{\delta p_a}+\frac{1}{3} s^a
\frac{\delta {\cal L}}{\delta p^0}+ d^{abc} s^b 
\frac{\delta {\cal L}}{\delta p_c}
\,\,\, ,
\label{ward}
\end{equation}
as well as the following correspondence of operators between
the heavy baryon and relativistic theories:
\begin{equation}
\bar{B}_v S_v^\mu \partial_\mu p B_v \leftrightarrow  -iM \bar{B} \gamma_5 p B \,\,\, ,
\label{correspondence}
\end{equation}
where $B$ and $B_v$ are the   relativistic and heavy baryon fields respectively.
 
The leading terms in the GTD are therefore of order $p^2$.
 There are several relations among the discrepancies that 
are exact  at LO.
One of them is the Dashen-Weinstein relation\cite{Dashen}:
\begin{equation}
m_K^2\, \left( \frac{g_A}{g_V} \right)^{NN\pi}\,\Delta^{NN\pi}=
\frac{1}{2}\;m_\pi^2\;
\left(3 \left( \frac{g_A}{g_V } \right)^{N\Lambda K} \Delta^{N\Lambda K}-
\left( \frac{g_A}{g_V}\right)^{N\Sigma K} \Delta^{N\Sigma K}\right).\label{DW}
\end{equation}
This particular relation provides useful insight as 
will be shown in the phenomenological discussion.

Since the bulk of the contribution to the GTD
 will result from the counterterms of Eq. (\ref{L3}), it is important to
 consider what physics determines their magnitude.
It seems   likely that a meson dominance model may provide the correct picture.
In such a model the size of the counterterms would be determined
by the lightest excited pseudoscalar
mesons that can attach the pseudoscalar current $\bar{q}\gamma_5 \lambda^a q$
to the baryons. The relevant such states are in the $\Pi'$ octet   consisting
 of
$\pi(1300)$, $\eta (1440)$ and $K(1460)$. The next set of pseudoscalar states
is in the range of 1800 to 2000 MeV, and thus, one may expect
that they only give corrections at the order of 20 to 30\%.
%(careful here since the
%$\eta(1440)$ might also need to be included; the question is who is the
%member of the octet: $\eta (1295)$ or $\eta(1440)$?.
%From Gell-Mann-Okubo relation:  $4 m_K-m_\pi-3 m_\eta=0$ it seems that the
%one that works best is the $\eta(1440)$).
 The meson dominance model
can be implemented using an effective Lagrangian in analogy with 
Ref. \cite{resonances}. The coupling of the $\Pi'$ octet to the
pseudoscalar current is obtained from the effective Lagrangian:
\begin{equation}
{\cal L}_{\Pi'} = \frac{1}{2} {\rm Tr}(\nabla^\mu \Pi' \nabla_\mu \Pi' )
- \frac{1}{2} M_{\Pi'}^2 \Pi'^2+i d_{\Pi'} {\rm Tr}(\Pi'\chi_{-})+... \,\,\, ,
\label{Lpi'}
\end{equation}
where we display only those terms relevant to our problem.
Here the $\Pi'$ octet responds to chiral rotations in the same way as the
baryon octet. The matrix element of
the divergence of the axial current is given by
\begin{eqnarray}
 <0\mid \partial^\mu A_\mu^a\mid\Pi'^b>&=&-\frac{B}{2}
 d_{\Pi'} {\rm Tr}(\lambda^b\{\lambda^a,{\cal M}_q\})\nonumber\\
 &=&-\delta^{ab} d_{\Pi'} m_a^2 ~~~ ,
\label{matrixelpi'}
\end{eqnarray}
 and the $\Pi'$-baryon coupling can be expressed through the effective Lagrangian:
\begin{eqnarray}
{\cal L}_{\Pi'B}&=&D' {\rm Tr}(\bar{B} \gamma_5 \{\Pi',\, B\})+
F' {\rm Tr}(\bar{B} \gamma_5 [\Pi',\, B]) \,\,\, .
\label{Lpi'baryon}
\end{eqnarray}
{}From Eqs. (\ref{matrixelpi'}) and (\ref{Lpi'baryon}) one readily obtains 
the   contribution to $\delta^{abc}$:
\begin{equation}
\delta^{abc}_{\Pi'}=
- d_{\Pi'} g^{abc}_{\Pi'B} \frac{m_c^2}{q^2-M_{\Pi'}^2} 
    \approx d_{\Pi'} g^{abc}_{\Pi'B} \frac{m_c^2}{M_{\Pi'}^2} \,\,\, .
\label{deltares}
\end{equation}
Here $g^{abc}_{\Pi'B}=\frac{F'}{\sqrt{8}} {\rm Tr}(\lambda^b [\lambda^c,\lambda^a])+
\frac{D'}{\sqrt{8}} {\rm Tr}(\lambda^b \{\lambda^c,\lambda^a\})$.
The current situation is that the couplings of the $\Pi '$ are not known,
and there is no estimate in the literature that one could judge reliable.
As we comment later, the GTD's actually serve to determine 
$d_{\Pi '}(q^2) g^{abc}_{\Pi 'B}$   much more precisely than any model calculation
available, provided the meson dominance model is realistic.

\section{Loop contributions}

%The first one-loop contributions that come to mind are those that affect
%the $q^2$ dependence of $g_{cab}$, and which are represented
%by the diagram in Fig. 2a.
There are several one-loop contributions to the GTD that we illustrate in Fig. 2.
There are also, at the same NLO,   two-loop contributions that we do not display here.
 Although we do not perform here a full calculation, we do arrive a some interesting
observations about such NLO effects by loops.  
Let us consider the loop diagram in Fig. 2a. We can show that in HBChPT this loop  
effect on the GTD is ${\cal O}(1/M^2)$, and must  therefore
be suppressed by two powers relative to the LO contribution.
Indeed, in HBChPT the diagram is proportional to the following loop integral:
\begin{equation}
-i T^{\mu \nu}\int\frac{d^dk}{(2 \pi)^d} \;\frac{k_\mu k_\nu}{k^2-m_d^2}
\;\frac{1+k \cdot v/(2M_f)+{\cal O}(1/M_f^2)}{k.v+k^2/(2M_f)-\delta m_{fa}}
\;\frac{1+(k+q).v/(2M_e)+{\cal O}(1/M_e^2)}{(k+q).v+(k+q)^2/(2M_e)-\delta m_{eb}} \,\,\, ,
\label{11}
\end{equation}
where $\delta m_{ab}\equiv M_a-M_b$, and $T^{\mu \nu}$ is transverse to the
four-velocity $v$. For spin 1/2 baryons in the loop 
$T^{\mu \nu}\propto S_v^\mu q{\cdot}S_v S_v^\nu$. It is also easy to
show explicitly that  $T^{\mu \nu}$ is transverse if one or both  
lines in the loop are spin 3/2 baryons. From energy-momentum
conservation we have
\begin{equation}
q.v=(M_b-M_a)-\frac{q^2}{2 M_b}.
\label{12}
\end{equation}
Using this and the transversity of $T^{\mu \nu}$,
 the expansion of Eq. (\ref{11}) shows no
$q^2$-dependence at ${\cal O}(1)$ and ${\cal O}(1/M)$. 
We conclude that the one-loop diagrams considered here 
must affect the GTD at ${\cal O}(1/M^2)$ \footnote{For a related
discussion, see Ref. \cite{Birse}.}, 
and are thus negligible in the large $M$ limit.
 Another type of one-loop contribution is not suppressed by $1/M$. 
 These  are the diagrams 
 involving the insertion of terms from ${\cal L}^{(3)}$ as shown in
Fig. 2b, which correct the GTD at NLO.  Similarly there are  
NLO  two-loop contributions that are of leading order in $1/M$.

It is interesting to comment    here on a one-loop calculation
in the framework of a relativistic baryon effective Lagrangian, as
used in Refs. \cite{Gasser,Goity}. It turns out that the relativistic
 version of the loop diagram in Fig. 2a gives a finite $q^2$ dependence
 to the $g_{cab}$ coupling, namely,

\begin{eqnarray}
g_{cab}(q^2)-g_{cab}(0)&=&
\left(
\frac{1}{2 F_\pi}\right)^3 \;
\sum_{d,e,f=1}^{8}
g_A^{afd}g_A^{ebd}g_A^{fec}
{\cal J}^{fed}(q^2, M_a,M_b,m_c)
%\nonumber\\
%g_A^{abc}&=& F {\rm Tr}(\lambda^a[\lambda^b,\lambda^c])+
%D {\rm Tr}(\lambda^a\{\lambda^b,\lambda^c\})
\label{9}
\end{eqnarray}
where 
%$F$ and $D$ are the beta-decay couplings at leading
% order in the chiral expansion, and 
 the integral
${\cal J}^{fed}$ is given by:
\begin{eqnarray}
 {\cal J}^{fed}(q^2, M_a,M_b,m_c) &=& {1\over{(4 \pi)^2}}
 {\cal C}(M_a,M_b,M_e,M_f)
 \int^1_0 dx \int^{1-x}_0 dy\;\; {{{\cal{N}}(x,y)}\over{{\cal{D}}(x,y)}} \,\,\, ,
 \label{10} \\
 2 {\cal{N}}(x,y) &=& (x+y-1)(M_a+M_b)^2-q^2(x+y)+2(1-x)M_aM_e
 +2xM_aM_f \nonumber \\
&+&2(1-y)M_fM_b+2yM_bM_e+(M_f-M_e)^2  ~~~,\\
{\cal{D}}(x,y) &=&
 (1-x-y) (x M_a^2 + y M_b^2 - M_d^2) - M_f^2 x - M_e^2 y + xy q^2 \,\,\, , \\
{\cal C}(M_a,M_b,M_e,M_f) &=& (M_b+M_e) (M_a+M_f) (M_e+M_f) \,\,\, .
\end{eqnarray} 
One can readily check that for SU(2) one obtains the result 
in Ref. \cite{Gasser}.

The interesting thing here is that the contribution to the GTD
 by the loop is not suppressed by $1/M$. Actually, it is nearly constant 
 for  
 baryon masses ranging from a few hundred MeV to an arbitrarily large mass.
This result seems at odds with the one from HBChPT, 
but the two can be harmonized
as follows: in the limit of large $M$ it turns out that in the relativistic
calculation there are contributions 
to the loop integral  from momenta that are ${\cal O}(M)$. $M$  
acts in fact as a regulator scale. In HBChPT on the other hand,  
   one is doing a $1/M$ expansion of the integrand, which  
 implies that one is assuming  a cutoff in the 
 loop integrals given by a QCD scale.
% Thus, in the large $M$ limit the relativistic calculation should be
% considered together with a counterterm that would subtract 
% the loop's leading (in 1/M) contribution
% and in this way bring agreement with
% the correct result directly obtained in HBChPT. 
The relativistic and HBChPT frameworks must each lead to the 
same physical results; in the present case this implies that
in order to lead to the same results for the discrepancies,
the coefficients $F_{19}$ and $D_{19}$ in ${\cal L}^{(3)}$
must be readjusted when going from one framework to the other.
In the real world, $M\sim \Lambda_\chi$
 and we may use  the relativistic calculation as an 
 estimate of this class of loop contributions to the discrepancy.
 For the discrepancies of interest herein, these loop 
contributions are small,
 between ten to twenty percent of the discrepancies themselves, 
 and smaller than their current errors. The numerical results are
 \begin{eqnarray}
\Delta_{loop}^{ N N\pi }&=&0.0043 \\
\Delta_{loop}^{ N \Lambda K }&=&-0.044 \\
\Delta_{loop}^{ N \Sigma K }&=&0.044,
\end{eqnarray}
where we use  $D=0.79$ and $F=0.46$ for the SU(3) axial vector couplings.

Of course the calculated loop contribution is not all that there is; the
inclusion of decuplet baryons in the loop also gives contributions to the
discrepancy. 
(Ref. \cite{GermCan} discusses some $\Delta(1232)$ effects with
only two quark flavors.)
Using Rarita-Schwinger propagators and three quark flavors, 
we have checked that the $q^2$-dependent part does show an UV divergence
in the relativistic framework.
HBChPT also permits two-loop contributions at NLO.
Currently a more complete calculation of the discrepancy 
at NLO is underway.\cite{GTD4}

\section{Results}

There are only three
discrepancies that can be determined from existing data
on baryon-pseudoscalar couplings: $\Delta^{NN\pi}$, 
$\Delta^{N\Lambda K}$, and $\Delta^{ N\Sigma K}$.

Due to the smallness of the u and d quark masses, 
$\Delta^{NN\pi}$ is necessarily 
very small, and its determination requires a very precise knowledge
of the $g_{\pi NN}$ coupling ($g_A$ and $F_\pi$ are already known to 
enough precision, leaving most of the uncertainty in the determination
of  $\Delta^{NN\pi}$ to the uncertainty in $g_{\pi NN}$). 
The most recent determination of $g_{\pi NN}$ from $NN$, $N\bar{N}$ and
$\pi N$ data is by the Nijmegen group\cite{Nij}. They analyzed a total of 
twelve thousand data and arrived at $g_{\pi NN}=13.05\pm 0.08$. 
Similar results are obtained by the
VPI group\cite{VPI}. Since the errors quoted are only statistical, in our 
fit below we will increase the error by about a factor of two
in order to roughly account for systematic uncertainties.
There is still some disagreement between determinations of $g_{\pi NN}$
 by different groups. 
Larger values have been obtained, such as    
$g_{\pi NN}=13.65\pm 0.30$ by Bugg and Machleidt\cite{Bugg}, 
and a similar result by Loiseau et al.\cite{Ericson}. As we find out below, 
 our analysis of the discrepancies strongly favors the smaller $g_{\pi NN}$
  values.
 Using
 $F_\pi=92.42$ MeV,  $\left(\frac{g_A}{g_V}\right)^{NN\pi}=-1.267\pm 0.004$ \cite{pdg},   
    Eq. (\ref{7}) gives,
\begin{eqnarray}\label{exptpinn1}
 \Delta^{NN\pi}_{\rm expt}&=&0.014\pm 0.006 {\rm ~~ ~for~~~} 
        g_{\pi NN}=13.05\pm 0.08, \\
 \Delta^{NN\pi}_{\rm expt}&=&0.056\pm 0.020 {\rm ~~~for~~~} 
        g_{\pi NN}=13.65\pm 0.30. \label{exptpinn2}
\end{eqnarray}
  
 The determination of the $g_{KN\Lambda}$ and  $g_{KN\Sigma}$ couplings
 relies on a  more sparse  data set. The Nijmegen group   analyzed 
 data from $Y\bar{Y}$ production at LEAR, and  
 they obtained\cite{Nij2}:  $g_{K N\Lambda}=13.7\pm 0.4$ and  
 $g_{K N \Sigma}=3.9\pm 0.7$. 
These values are consistent with an earlier analysis by Martin\cite{Martin},
 where only an upper bound for $g_{K N \Sigma}$ is given.
 Using $F_K=1.22\; F_\pi$ and  
 $\left( \frac{g_A}{g_V}\right) ^{N\Lambda K}=-0.718\pm 0.015$ and $\left(\frac{g_A}{g_V}\right)^{N\Sigma K}=0.340\pm 0.017$  \cite{pdg},
 Eq. (\ref{7}) gives,
\begin{eqnarray}
 \Delta^{N\Lambda K }_{\rm expt} &=& 0.17\pm 0.03 \\
 \Delta^{N\Sigma K }_{\rm expt} &=& 0.17\pm 0.14
\end{eqnarray}
 
Disregarding SU(2) breaking, which implies that there is no contribution
from the term proportional to $b_{20}$ to these discrepancies,
we can use the three measured discrepancies to
determine the two LO parameters in HBChPT:
\begin{eqnarray}\label{F}
M\,F_{19} &=& 0.4 \pm 0.1\;{\rm GeV^{-1}}, \\
M\,D_{19} &=& 0.7\pm 0.2\;{\rm GeV^{-1}},
\label{D}
\end{eqnarray}
where $M$ is here the common baryon-octet mass in the chiral limit. 
Both choices for $g_{\pi{N}N}$,
given in Eqs.
(\ref{exptpinn1}) and (\ref{exptpinn2}), lead to values for
$M\,F_{19}$ and $M\,D_{19}$ that agree within the quoted
uncertainties.  The LO discrepancies resulting from our fit are:
\begin{eqnarray}
\Delta^{NN\pi} &=& 0.017; \;\; 0.018, \\
\Delta^{N\Lambda K } &=& 0.17; \;\; 0.18, \\
\Delta^{N\Sigma K } &=& 0.17; \;\; 0.19,
\end{eqnarray}
where the quoted results correspond respectively to the smaller and larger $g_{\pi N N}$ couplings.
The larger value $\Delta^{NN\pi}= 0.056$ of Eq. (\ref{exptpinn2}), 
corresponding to the larger 
$g_{\pi NN}$ coupling, cannot come out consistently from the fit.  
To understand this one can use the Dashen-Weinstein relation, Eq. (\ref{DW}),
which holds exactly in our LO calculation. 
For the  results of  the discrepancies involving the hyperons
   the term proportional to  
$\Delta^{N\Sigma K }$ in the Dashen-Weinstein relation is about one fifth of that proportional to
$\Delta^{N\Lambda K }$, and the right hand side of Eq. (\ref{DW})
 would imply  that  $\Delta^{NN\pi}$ must  be   about $1.5\%$.
The only way to accomodate a larger $\Delta^{NN\pi}$ would be  larger 
$\Delta^{N\Lambda K }$ and  $\Delta^{N\Sigma K }$ or else
a large deviation from the Dashen-Weinstein relation. The latter seems 
unlikely because the corrections to the
relation must be suppressed by two powers in HBChPT
(this is so because the corrections to 
the axial-vector couplings and to the  discrepancies are of  ${\cal O}(p^2)$).
On the other hand the former possibility would require that the magnitudes
of $g_{K N\Lambda}$ and  $g_{K N\Sigma}$
be unrealistically large. In fact, $\Delta^{N\Lambda K }$ and $\Delta^{N\Sigma K }$
should be close to  unity, implying a serious failure of the
low energy expansion. Thus, we conclude that only the smaller values 
of $\Delta^{NN\pi}$, and thus of $g_{\pi NN}$,   
are consistent. This shows the importance of the current analysis of the GTD in SU(3).

Finally, the coupling constants required in the meson dominance model
resulting from our analysis are as follows:
\begin{eqnarray}
d_{\Pi'} F' &=& 2.4\pm 0.5 {\rm GeV} \\
d_{\Pi'} D' &=& 4.5\pm 0.5 {\rm GeV}.
\end{eqnarray}
Since here $F'$ and $D'$ are baryon-meson couplings, it is not 
unreasonable that they should have values similar to those of, say, 
the pion-nucleon coupling. This would imply that the coupling $d_{\Pi'}$ 
should be a few hundred MeV.
This makes the meson dominance picture quite plausible.

In conclusion, we have shown that the GTD in SU(3) is given at leading order by
two tree-level contributions, and 
that the corrections are suppressed by two powers in HBChPT.
Some of the loop corrections were calculated explicitly and found to be small.
Our leading order analysis indicates a strong preference for a smaller 
Goldberger-Treiman discrepancy
in the pion-nucleon sector, thus favoring the smaller values of the 
pion-nucleon coupling extracted in  recent partial wave analyses.
 \vfill
\newpage
 
\section*{Acknowledgements}
We would like to thank Juerg Gasser for allowing us to use
material from an earlier unpublished collaboration and for useful 
discussions. We
also thank G. H\"ohler and Ulf Mei\ss ner for useful comments, and 
Jan Stern for bringing to our attention  the Dashen-Weinstein relation.
This work was
supported by the National Science Foundation through grant \# HRD-9633750
(JLG and MS), and \# PHY-9733343 (JLG) and by the Department of Energy
through contract DE-AC05-84ER40150 (JLG, RL), and in part by Natural
Sciences and Engineering Research Council of Canada (RL), the
Fundaci\'on Antorchas of Argentina (MS) and by the grant \# PMT-PICT0079
of the ANPCYT of Argentina (MS).

\vfill
\eject

\begin{figure}
\epsffile{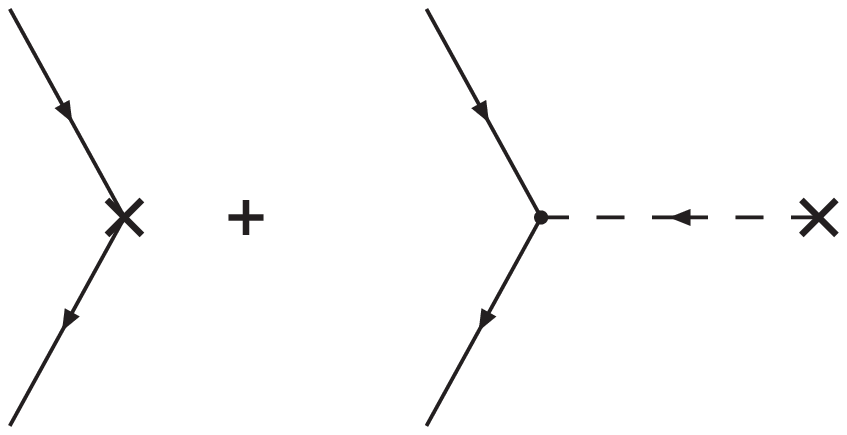}
\caption{Diagrams representing the contact term and the pole term
in the matrix elements of the divergence of the axial currents. 
Crosses represent the divergence of axial currents.}
\end{figure}

\newpage

\begin{figure}
\epsffile{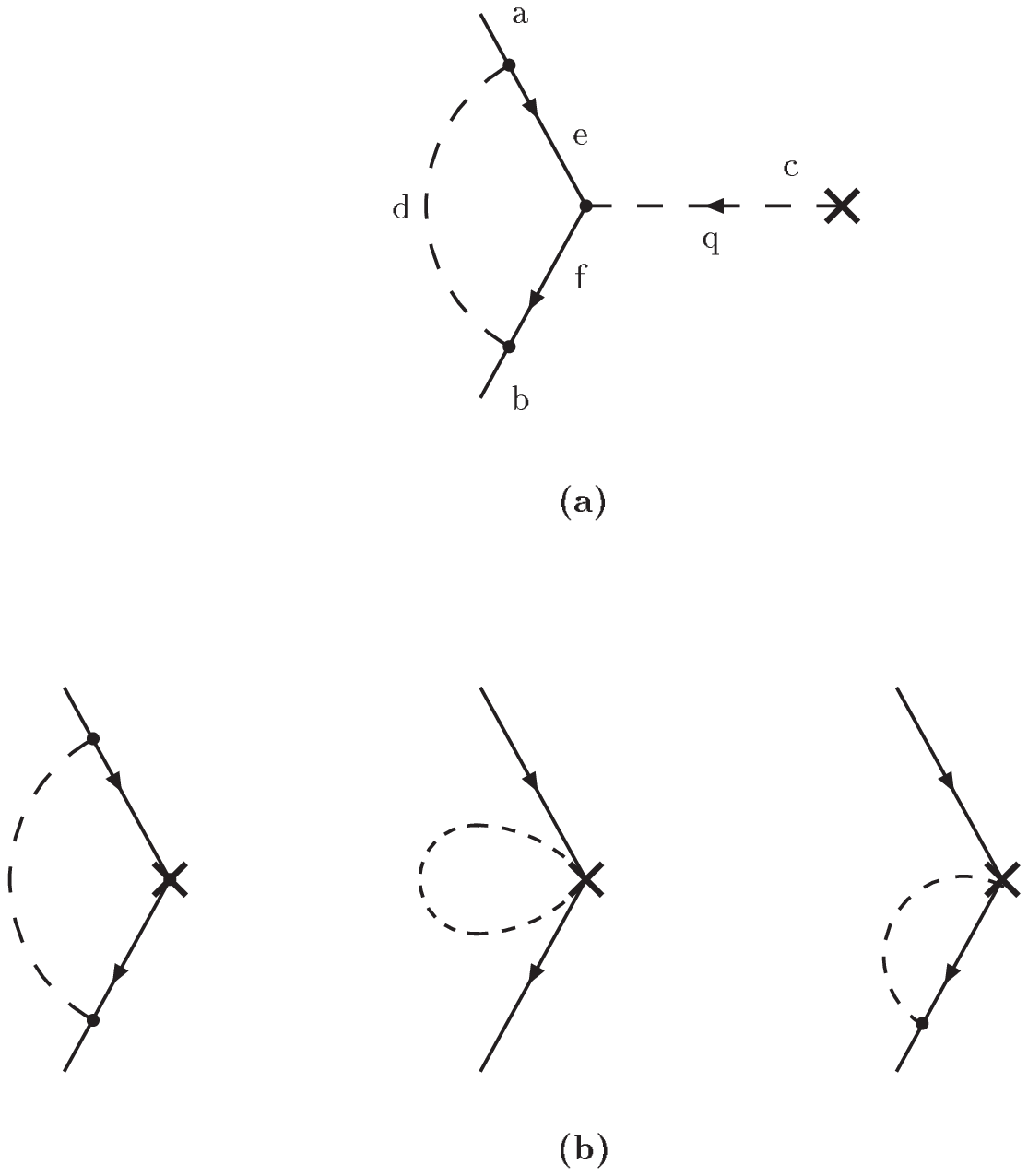}
\caption{One loop diagrams that give NLO corrections to the GTD. 
In (a) the cross represents the divergence of the axial current
obtained from the ${\cal O}(p^2)$ Lagrangian, and in (b) the same divergence 
obtained from the ${\cal O}(p^3)$ baryon Lagrangian.}
\end{figure}


\newpage

\begin{thebibliography}{99}

\bibitem{GT}
M. L. Goldberger and  S. B. Treiman, Phys. Rev. {\bf 110} (1958) 1178.
\bibitem{Dominguez} C. A. Dominguez, 
Riv. Nuovo Cim. {\bf 8} (1985) 1, and references therein.
\bibitem{Gasser} J. Gasser, M. E. Sainio and A \v{S}varc, Nucl. Phys. 
{\bf B307} (1988) 779.
\bibitem{HBChPT} E. Jenkins, A. V. Manohar, Phys.Lett. {\bf B255} 
(1991) 558.
\bibitem{HUGS} U-G. Meissner, in \lq\lq Themes in Strong Interactions",
Proceedings of the $12^{th}$ HUGS at CEBAF,
J. L. Goity Editor, World Scientific (1998) 139, and references therein.
\bibitem{Stern} N. H. Fuchs, H. Sazdjian and J. Stern, Phys. Lett. {\bf B 238} (1990) 380.
\bibitem{Dashen} R. Dashen and M. Weinstein Phys. Rev. {\bf 188} (1969) 2330.
\bibitem{Ecker} G. Ecker and  M. Moj\v{z}i\v{s},  
Phys. Lett. {\bf B365} (1996)  312. 
\bibitem{bugfix} N. Fettes, U.-G. Meissner and S. Steininger,
Nucl. Phys. A640, 199 (1998).
\bibitem{Lewis} H. W. Fearing, R. Lewis, N. Mobed and S. Scherer,
 Phys. Rev. D56, 
(1997) 1783.
\bibitem{resonances} G. Ecker, J. Gasser, A. Pich, and E. de Rafael,
 Nucl. Phys. {\bf B321} (1989)  311.
\bibitem{Birse} J. A. McGovern and  M. C. Birse, MC-TH-98-13 preprint (1998).
e-Print Archive: hep-ph/9807384. 
\bibitem{Goity} J. Gasser and J. L. Goity, unpublished.
\bibitem{GermCan} V. Bernard, H. W. Fearing, T. R. Hemmert and U.-G. Meissner,
Nucl. Phys. A 635, 121 (1998); Nucl. Phys. A 642, 563 (1998).
\bibitem{GTD4} J. L. Goity, R. Lewis, M. Schvellinger and L. Zhang, in progress.
\bibitem{Nij}
J.J. de Swart, M.C.M. Rentmeester and R.G.E. Timmermans, Proc. of MENU97, TRIUMF Report TRI-97-1 (1997) 96.
\bibitem{VPI} R. A. Arndt, I. I. Strokovsky and R. L. Workman, Phys. Rev. {\bf C 52}
(1995) 2246.
\bibitem{Bugg}
D.V. Bugg and R. Machleidt, Phys. Rev. {\bf C 52} (1995) 1203.
\bibitem{Ericson} B. Loiseau, T.E.O. Ericson, 
J. Rahm, J. Blomgren and N. Olsson, $\pi$N-Newsletter {\bf 13} (1997) 117.
\bibitem{pdg}
Particle Data Group (C. Caso {\it et al.}), Eur.\ Phys.\
J. C {\bf 3}, (1998) 1.
\bibitem{Nij2} R. G. E. Timmermans, T. A. Rijken and J. J. de Swart, Phys. Lett. {B 257} (1991) 227.\\
R.E.G. Timmermans, Th. A. Rijken and J.J. de Swart, Nucl. Phys. {\bf A 585} (1995) 143c.
\bibitem{Martin} A. D. Martin, Nucl. Phys. {\bf B 179} (1981) 33.
\end{thebibliography}
\end{document}